\documentclass[iop]{emulateapj}

\shorttitle{The ultra-long GRB 111209A}
\shortauthors{Stratta et al.}

\begin{document}


\title{The ultra-long GRB 111209A \\
       II. Prompt to afterglow and afterglow properties}


\author{G. Stratta}
\affil{Osservatorio Astronomico di Roma (OAR/INAF), via Frascati 33, 00040, 
       Monte Porzio Catone, Italy}

\author{B. Gendre}
\affil{ARTEMIS, UMR 7250 (CNRS/OCA/UNS), boulevard de l'Observatoire, BP 4229, F 06304 Nice Cedex, France}

\author{J.L. Atteia}
\affil{Universit\'e de Toulouse, UPS-OMP, IRAP, Toulouse, France}
\affil{CNRS, IRAP, 14, avenue Edouard Belin, F-31400 Toulouse, France}

\author{M. Bo\"{e}r}
\affil{ARTEMIS, UMR 7250 (CNRS/OCA/UNS), boulevard de l'Observatoire, BP 4229, F 06304 Nice Cedex, France}

\author{D. M. Coward}
\affil{School of Physics, University of Western Australia (UWA), Crawley WA 6009, Australia}


\author{M. De Pasquale}
\affil{Mullard Space Science Laboratory (MSSL), University College London, Holmbury St. Mary, Dorking Surrey, RH5 6NT, UK}

\author{E. Howell}
\affil{School of Physics, University of Western Australia (UWA), Crawley WA 6009, Australia}

\author{A. Klotz}
\affil{IRAP, 14, avenue Edouard Belin, F-31400 Toulouse, France}

\author{S. Oates}
\affil{Mullard Space Science Laboratory (MSSL), University College London, Holmbury St. Mary, Dorking Surrey, RH5 6NT, UK}

\and

\author{L. Piro}
\affil{Istituto di Astrofisica e Planetologia Spaziali di Roma (IAPS/INAF), via fosso del cavaliere 100, 00133 Roma, Italy}




\begin{abstract}
The ``ultra-long" Gamma Ray Burst GRB 111209A at redshift z=0.677, is so far the longest GRB ever observed, with rest frame prompt emission duration of $\sim 4$ hours. In order to explain the burst exceptional longevity, a low metallicity blue supergiant progenitor has been invoked. In this work, we further constrain the phenomenology and progenitor properties of this peculiar GRB by performing a multi-band temporal and spectral analysis of both the prompt and the afterglow emission. We use proprietary and publicly available data from Swift, Konus Wind, XMM-Newton, TAROT as well as from other ground based optical and radio telescopes. We find some peculiar properties that are possibly connected to the exceptional nature of this burst, namely: 
i) an unprecedented large optical delay of $410\pm50$ s between the peak time in gamma-rays and the peak time in optical of a marked multiwavelength flare;
ii) multiwavelength prompt emission spectral modelling requires a certain amount of dust in the circumburst environment, with rest frame visual dust extinction of $A_V=0.3-1.5$ mag, that may undergo to destruction at late times;  
iii) we detect the presence of a hard spectral extra power law component at the end of the X-ray “steep decay phase” and before the start of the X-ray afterglow, which was never revealed so far in past GRBs.
The optical afterglow shows more usual properties, with a flux power law decay with index $1.6\pm0.1$ and a late re-brightening feature observed at $\sim1.1$ day after the first BAT trigger. We discuss our findings in the context of several possible  interpretations given so far to the complex multi-band GRB phenomenology 
and propose a binary channel formation for the blue supergiant progenitor. 

\end{abstract}





\section{Introduction}
\label{s:intro}

Long Gamma Ray Bursts are commonly interpreted in the context of the collapsar model \citep{Woosley1993,Paczynski1998,MacFadyen1999,Woosley2006}, where a massive, highly rotating star collapses into a black hole or a neutron star forming a temporary torus of matter around the central object. In this scenario, the accretion onto the central object is the engine that produces a burst of radiation expected to last from few seconds to few tens of seconds. The typical durations of GRB prompt emission at high-energy have a distribution that peaks at 20-30 seconds, with a range that goes from few seconds up to hundreds of seconds  \citep{Kouveliotou1993}, consistent with the expected time scale of a gravitational collapse of the core of a massive star as the CO Wolf-Rayet (WR) star, as predicted by the collapsar model. 

GRB 111209A was an exceptional ``ultra-long" GRB at redshift z=0.677 \citep{Vreeswijk2011} with an unprecedented burst duration of few hours \citep{Hoversten2011a}. The existence of such ultra-long bursts as GRB 111209A, together with other very long GRBs as for example GRB 101225 and GRB 121027A \citep[e.g.][]{Levan2013}, has imposed some modifications to the standard collapsar model mentioned above. Among the scenarios proposed to explain these events, possible solutions are a more massive and extended progenitor star \citep[e.g.][]{Gendre2013,Nakauchi2013}, or fall-back accretion of a partly survived progenitor envelope \citep[e.g.][]{Wu2013,Quataert2012}.

Motivated by the unique properties of the prompt emission of GRB 111209A, we proposed (Gendre et al. 2013; hereafter Paper I) that this burst could originate from a blue supergiant, an hypothesis that has been further investigated by \cite{Nakauchi2013}. 
Indeed, following \cite{Woosley2012}, in Paper I we found that the long duration and high luminosity of GRB 111209A in gamma-rays, and the lack of evidence of a bright supernova during the successful follow-up campaign up to dozens of days after the trigger despite the rather low redshift of this burst (see Levan et al. 2013), favor a rare type of low-metallicity collapsing blue supergiant as progenitor, making bursts like GRB 111209A a better prototype of the expected Pop III stars than normal long GRBs \citep[e.g.][]{Kashiyama2013}.  
GRB 111209A cannot be considered as an extreme representative of the long GRB class, and we showed evidence in Paper I for which we had to invoke a different progenitor. This is also the conclusion reached later by \cite{Levan2013}. 

In this work we investigate the global properties of GRB 111209A, taking advantage of the extended data set we could acquire both during the prompt phase and the afterglow. 
We show that, while there are some similarities of the multiwavelength afterglow emission with ``normally" long GRBs, the prompt emission and the prompt-to-afterglow transition present peculiar properties. The overall picture emerging from our analysis supports the evidence of a non negligible metallicity in the host galaxy \citep{Levan2013}. We tentatively suggest a binary system progenitor for GRB 111209A to deal with the metal-enriched environment.

In the following we present the data set we used in our analysis (\S \ref{s:datared}), the temporal and spectral properties of the prompt emission (\S \ref{s:prompt}) and of the early and late afterglow (\S \ref{s:afterglow}). We then discuss our findings (\S \ref{s:discussion}) and show our conclusions (\S \ref{s:conclusions}). 

Throughout the paper, we describe the temporal and spectral power law indices according to the notation $F(\nu,t) \propto \nu^{-\beta} t^{-\alpha}$, unless otherwise specified.

\section{The data}
\label{s:datared}

   \begin{figure*}
   \centering
   \includegraphics[width=13cm]{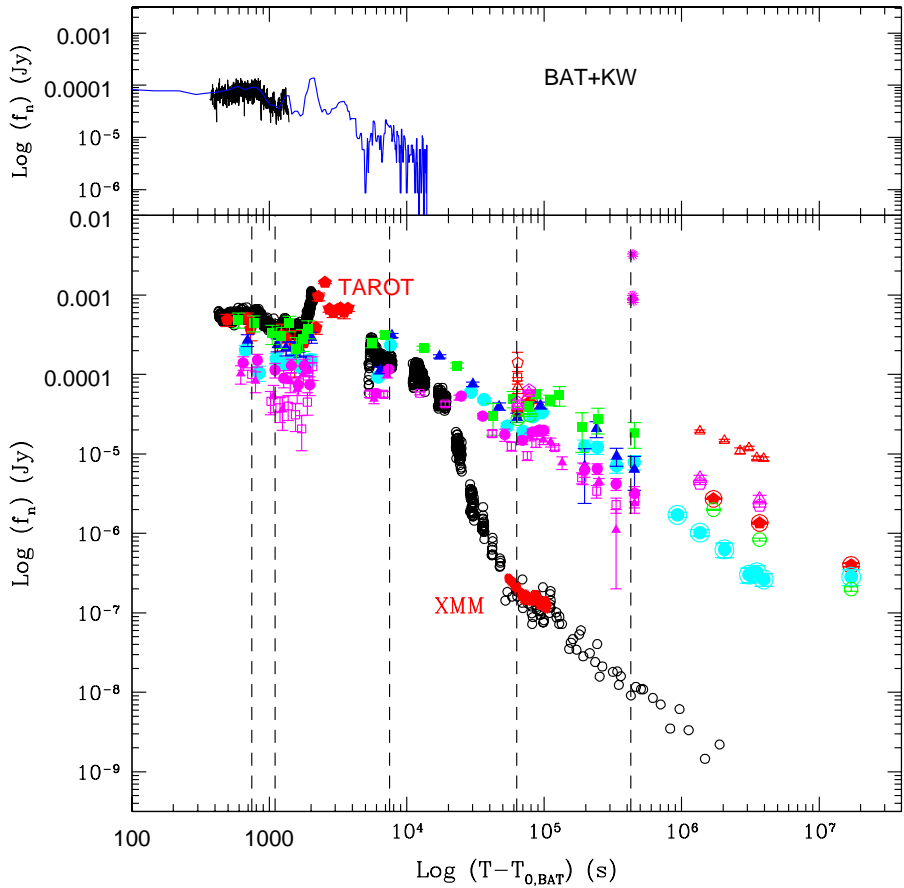}
   \caption{The multi-wavelength light curve of GRB 111209A. Top panel: {\it Swift}/BAT (black) and Konus-Wind (blue). Bottom panel: {\it Swift}/XRT (black open circles), XMM (starred red circles), TAROT (filled red penthagons), {\it Swift}/UVOT: {\it w2}-band (magenta open squares), {\it m2}-band (magenta filled triangles), {\it w1}-band (magenta filled circles), {\it u}-band (cyan filled circles), {\it b}-band (blue filled triangles), {\it v}-band (green filled squares), and ground based telescopes from Levan et al. (2013) and GCN Circulars: u-band (cyan dotted circles), g-band (green open circles) r-band (red encircled penthagons), i-band (magenta open triangles), z-band (magenta open penthagons), J-band (red open triangles), H-band (red open squares), K-band (red open penthagons), radio-band (magenta stars).  The vertical dashed lines indicate the central epochs at which we extract radio/optical to X-ray spectral energy distribution (see text): $650-850$, $1-1.2$ ks and $7-8$ ks (prompt emission), $62-64$ ks (GROND data epoch) and $426-466$ ks after the trigger (radio data epoch).}
              \label{figure:all}%
    \end{figure*}

The Burst Alert Telescope (BAT) coded mask telescope on board {\it Swift} was triggered twice by GRB 111209A on the 9th of December 2011, at $T_0=07:12:08$ UT and on $T_0+424$s (trigger numbers 509336 and 509337). The {\it Swift} X-ray  telescope (XRT) monitoring started $T_0+419$ s and the UV-optical telescope (UVOT) began settled observations at 427 s post trigger \citep{Hoversten2011b}. Simultaneous R-band observations were performed with the TAROT Calern telescope \citep{Klotz2009} from $T_0+492$ s up to $T_0+3.7$ ks \citep{Klotz2011}. A successful ToO has been performed with XMM-Newton \citep{Gendre2011} for a total exposure time of 51.5 ks between $T_0+56.7$ ks and $T_0+108.2$ ks. 
This GRB was detected also with Konus-Wind at $T_0-1900$ s with evidence of a weaker emission from $T_0-5400$ s to $T_0-2600$ s  \citep{Golenetskii2011}. Light curves from Konus-Wind (KW) have been publicly released in the GRB Coordinates Network Circular, and are accessible through the web\footnote{http://www.ioffe.rssi.ru/LEA/GRBs/GRB111209A}. 

We refer the reader to Paper I for the methods of reduction of the data from {\it Swift}/XRT as well as from XMM-Newton.   
To convert the KW counts into flux density we first computed the background subtracted total counts from the digitilized light curve in the 20-1400 keV energy range during the main burst, that is from $T_0$-1990 s to $T_0$+4400 s, where we know the fluence to be of $(4.86 \pm 0.61)\times10^{-4}$ erg cm$^{-2}$ in the energy range $20$ keV $<E<1400$ keV \citep{Golenetskii2011}. By assuming the claimed best fit spectral model, that is a cut-off power law model $dN/dE \propto (E^{\Gamma_{KW}})e^{-E(2+\Gamma_{KW})/E_{cut}}$ with best fit photon index  $\Gamma_{KW}=-1.31\pm 0.09$ and cut-off energy $E_{cut} =310\pm53$ keV, we could estimate the flux density at the mean energy of the 20-1400 keV photon spectrum, that is at $\sim 116$ keV. 

NIR and optical afterglow data from the GROND telescope as well as from Gemini and VLT telescopes, have been taken from \cite{Kann2011a} and from \cite{Levan2013} and corrected for the Galactic extinction $E(B-V) = 0.02$ mag in the direction of the burst \citep{Schlegel1998}. Radio fluxes of GRB 111209A simultaneously measured with the Australia Telescope Compact Array (ATCA) in two bands, 5.5 and 9 GHz, at the mean observing time of 11:12UT, that is at $T_0+446.4$ ks or $T_0+5.2$ days ($\sim3.5$ days after optical re-brightening peak) have been taken from \cite{Hancock2011}. 

The resulting multiwavelength light curve of GRB 111209A is plotted in Figure \ref{figure:all}. 

\subsection{{\it Swift}/BAT data reduction}
To build the {\it Swift}/BAT 15-150 keV flux density light curve, we convert fluxes taken from the light curve repository\footnote{http://www.swift.ac.uk/burst\_analyser/} \citep{Evans2009,Evans2010} into flux densities assuming a power law model and using the available time resolved photon index estimates. We computed the flux density at the energy of $\sim 47$ keV, that is at the mean energy of the 15-150 keV average photon spectrum \citep{Palmer2011}. 

We also extract the energy spectra at two epochs where simultaneous UVOT and XRT data were available, in order to perform multi-band analysis. To this purpose, data (obsid 00509337 000) were reduced following the standard procedure described in the BAT data analysis threads\footnote{http://heasarc.nasa.gov/docs/{\it Swift}/analysis/}. In order to estimate the proper geometric parameters to make accurate response matrix, we use the task \texttt{batmaskwtevt} to build the missing auxiliary raytracing file, where we adopt the optical {\it Swift}/UVOT coordinates RA=14.34429 deg and Dec=-46.80106 deg \citep{Hoversten2011b}. 

\subsection{{\it Swift}/UVOT data reduction}

The optical counterpart of GRB 111209A was detected in all 7 UVOT filters. Observations were taken in both image and event modes. Before extracting count rates from the event lists, the astrometry was refined following the methodology described in \cite{Oates2009}. For both the event and image mode data, the source counts were extracted using a region of 5\arcsec or 3\arcsec radius. As it is more accurate to use smaller source apertures when the count rate is low \citep{Poole2008}, the 3\arcsec aperture was used when the count rate fell below 0.5 counts s$^{-1}$. In order to be consistent with the UVOT calibration, these count rates were then corrected to 5\arcsec using the curve of growth contained in the calibration files. Background counts were extracted using a circular region of radius 20\arcsec from a blank area of sky situated near to the source position. The count rates were obtained from the event and image lists using the {\it Swift} tools \texttt{uvotevtlc} and \texttt{uvotsource}, respectively. They were converted to magnitudes using the UVOT photometric zero points \citep{Breeveld2011}. The analysis pipeline used software HEADAS 6.10 and UVOT calibration 20130118.

\begin{table}
\tabletypesize{\scriptsize}
\tablewidth{0pt}
\caption{TAROT R-band observations (see \S\ref{s:tarot}). Flux and apparent magnitudes are not corrected for Galactic extinction.  }             
\label{t:tarot}      
\centering                          
\begin{tabular}{  c c c c c }        
\hline\hline                 
  $T_{start}$  & $T_{stop}$    &   Flux  & mag  & dmag  \\
  min	&	min & mJy  &  AB & AB\\
\hline
     7.95 &     8.45 &   0.4875 &   17.18 &  0.21 \\  
    11.20 &    11.70 &   0.5105 &   17.13 &  0.20   \\
    11.87 &    12.37 &   0.3733 &   17.47 &  0.27   \\
    19.61 &    21.11 &   0.3133 &   17.66 &  0.29   \\
    21.28 &    22.78 &   0.2884 &   17.75 &  0.33   \\
    24.62 &    26.12 &   0.2249 &   18.02 &  0.33   \\
    28.86 &    30.36 &   0.2512 &   17.90 &  0.28   \\
    35.53 &    37.03 &   0.3873 &   17.43 &  0.17   \\
    37.20 &    38.70 &   0.955 &   16.45 &  0.26   \\
    40.80 &    43.80 &   1.445 &   16.00 &  0.13   \\
    43.96 &    46.96 &   0.6668 &   16.84 &  0.15   \\
    47.13 &    50.13 &   0.6026 &   16.95 &  0.21   \\
    50.30 &    53.30 &   0.6546 &   16.86 &  0.21   \\
    53.46 &    56.46 &   0.6855 &   16.81 &  0.15   \\
    56.63 &    59.63 &   0.6138 &   16.93 &  0.27   \\
    60.61 &    63.61 &   0.6792 &   16.82 &  0.15   \\
  1037.91 &  1311.70 &   0.0198 &   20.66 &  0.10  \\
\hline
\end{tabular}
\end{table}

\subsection{TAROT data reduction}
\label{s:tarot}
Exposure of the field of GRB~111209A with the TAROT Calern robotic telescope \citep{Klotz2009} was taken with the tracking speed adapted to obtain a small ten pixel trail. This technique was used in order to obtain continuous temporal information during the exposure \citep{Klotz2006}. The spatial sampling was 3.29\arcsec pix$^{-1}$ and the FWHM of stars (in the perpendicular direction of the trail) was 2.05 pixels. Only the first exposure was performed with this technique. Successive images were tracked on the diurnal motion using exposure times increasing from 30 s to 180 s. 

Images were not filtered. We used the star NOMAD-1 0431-0011481 (R=11.56, V-R=+0.34) as a constant template for calibration for all TAROT images. We used the AudeLA software\footnote{http://www.audela.org/} to compute the magnitude using a Point Spread Function fit with the template star. The fit leads to the optical flux ratio GRB / template which is converted into R magnitudes. Results are quoted in Table \ref{t:tarot}.

   \begin{figure}
   \centering
   \includegraphics[width=9cm]{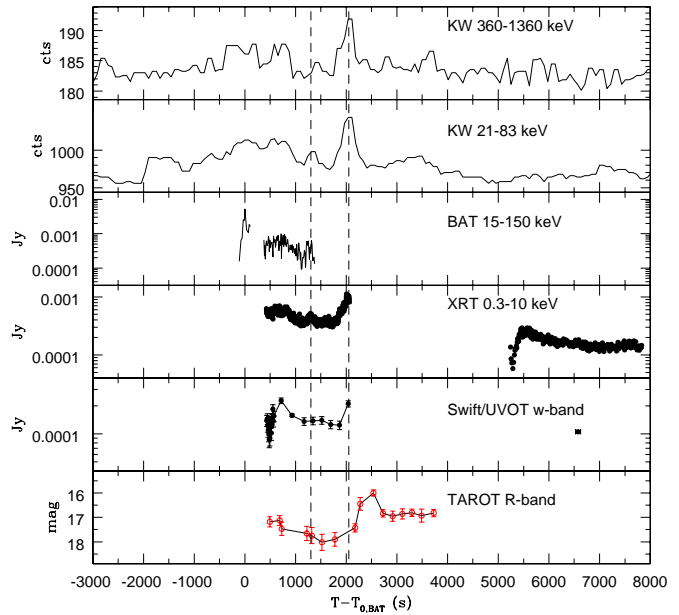}
   \caption{Prompt emission light curve between -3000 and 8000 s from the BAT trigger, at different wavelengths. Vertical dashed lines indicate the peak of two high energy flares computed at $T_0+1.3$ ks from {\it Swift}/XRT data analysis and at $T_0+2.0$ ks from Konus-Wind light data analysis. }
              \label{figure:promptlc}%
    \end{figure}

\section{The prompt emission}
\label{s:prompt}

In this section we analyze the prompt emission temporal and spectral properties by using simultaneous optical, X-ray and gamma-ray data.

\subsection{The optical flare temporal lag}
\label{s:lag}

In Figure \ref{figure:promptlc} we plot a light curve comparison in each energy band for the time interval $T_0-3000$ s to $T_0+8000$ s, where $T_0$ refers to the first BAT trigger time (see \S \ref{s:datared}).  Between $\sim T_0+400$ s and $\sim T_0+4000$ s the prompt emission of GRB 111209A was observed in a large frequency window, that is from optical wavelengths to gamma-rays. 

A pronounced flare is observed in the KW data at $\sim T_0+2$ ks and clearly visible at all wavelengths, although a complete monitoring before and after the flare peak epoch was possible only with the TAROT R-band and KW data. Assuming a gaussian, we could measure a peak epoch at $(2460\pm50)$ s after $T_0$ and width of $\sim130$ s using the TAROT data, while using KW data we measure a peak epoch at $(2050\pm10)$ s after $T_0$ and a width consistent with the optical one. The delay of the R band peak epoch is thus of $410\pm50$ s in respect with the gamma-ray peak epoch, that is, $\sim245$ s in the rest frame of the burst. 

We discuss some possible origins of this temporal lag in \S \ref{s:lagdisc}.

   \begin{figure*}
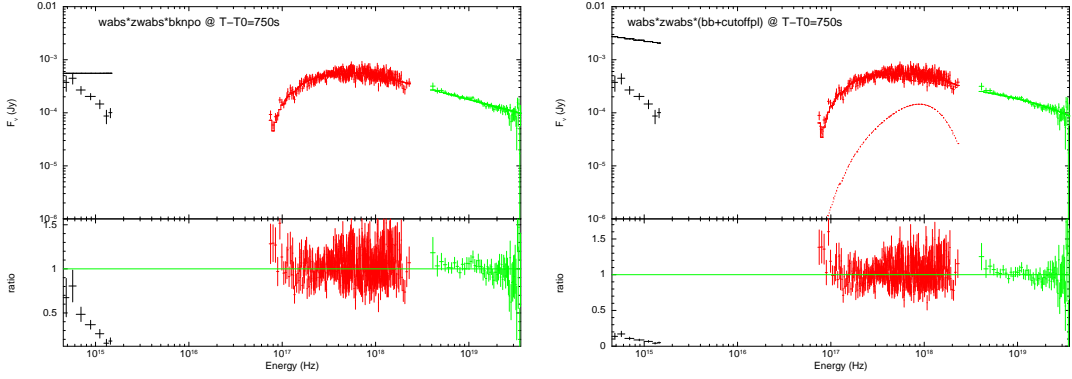

   \centering
\includegraphics[width=5cm,angle=-90]{fig3a_new.eps}
\includegraphics[width=5cm,angle=-90]{fig3b_new.eps}
   \caption{Prompt emission X-ray to optical SED centered at 750 s after the trigger where simultaneous TAROT (R band) and {\it Swift}/UVOT {\it v, b, u, w1, m2, w2} bands (black), XRT 0.3-10 keV (red) and 15-150 keV BAT (green) data are available. The black lines are the best fit continuum obtained from the simultaneous fit of the BAT and XRT data by assuming a broken power law model and a cut-off power law model plus a black body component. The optical fluxes are systematically below the model and can be fit assuming a significant dust extinction. In Table 2 we quote the results obtained by including the optical data in the fit.} 
              \label{figure:promptsed}%
    \end{figure*}

\begin{table*}
\tabletypesize{\scriptsize}
\tablewidth{0pt}
\caption{Multi-band prompt spectral analysis at three different epochs. Best fit parameters for each spectral model are quoted in three rows where the first and the second rows refer to the periods 650-850s and 1000-1200 s after $T_0$ where simultaneous TAROT, {\it Swift}/UVOT, {\it Swift}/XRT and {\it Swift}/BAT were available, while the third row refers to the preiod 7-8 ks after $T_0$ where {\it Swift}/XRT and UVOT data were available.  Three different spectral continuum were assumed (BPL=broken power law, Band=Band model, CPL=cut-off power law) plus a black body component (BB). The $\chi^2$ of the same fit without the BB component is also quoted ($\chi^2_{no BB}$) as well as its statistical significance computed through an F-test. The optical dust reddening was estimated in the rest frame of the burst and is best fit with a SMC-like extinction curve, apart for the CPL model during the first two epochs for which a MW extinction law better fit the data.  }             
\label{table:prompt}      
\centering                          
\begin{tabular}{ c c c c c c c c | c c}        
\hline\hline                 
$T-T_0$& $\beta_1$ & $\beta_2$ & $E_0$ & $N_{H,z}$ & $E(B-V)$ & T & $\chi^2$(d.o.f.) & $\chi^2_{no BB}$(d.o.f.)  & F-test\\    
ks &          &                 & keV    & $10^{21}$cm$^{-2}$ & mag & keV & & & p-value\\
\hline                        

&BPL+BB        &           &  				     & &  & &  & & \\
0.65-0.85&$0.0\pm0.1$ & $0.47\pm0.02$ & $3.7^{+0.8}_{-0.4}$ & $3.3\pm0.4$ & $0.17\pm0.03$ & $0.04^{+0.02}_{-0.01}$ & 483(446)   & 501(448) & 3e-4   \\
1-1.2&$0.1\pm0.1$ & $0.57\pm0.02$ & $2.7\pm0.2$ & $2.5\pm0.2$ & $0.24\pm0.03$ & $0.04\pm0.01$ & 401(414)     & 420(416) & 7e-5 \\     
7-8& $-0.7\pm0.6$ & $0.7\pm0.2$ & $0.02^{+0.60}_{-0.01}$ & $2.9\pm0.2$ & $0.22\pm0.03$ & $1.25\pm0.2$ & 262(288)  & 283(290) & 1e-5 \\     
\hline
&Band+BB             &            & &     &  & &  &  & \\
0.65-0.85& $0.0\pm0.1$ & $0.49\pm0.02$ & $16^{+7}_{-4}$ & $3.7\pm0.4$ & $0.14\pm0.03$ & $0.05\pm0.01$ & 473(466)  & 498(448) & 3e-4  \\    
1-1.2 & $0.0\pm0.1$ & $0.59\pm0.02$ & $9\pm2$ & $3.0\pm0.3$ & $0.23\pm0.03$ & $0.04\pm0.01$ & 393(414)   &  422(416)& 4e-7 \\   
7-8&$0.11\pm0.03$ & $1.7\pm0.2$ & $1.8^{+0.9}_{-0.5}$ & $2.4\pm0.2$ & $0.26\pm0.03$  & $1.25\pm0.2$ & 273(288)    & 286(290)& 1e-3 \\ 
\hline   
&CPL +BB                        &               &  &     & &  & &  &  \\
0.65-0.85& $0.25\pm0.01$ & $-$ & $\leq257$ & $2.9\pm0.3$ & $0.35\pm0.02$ & $1.3\pm0.1$ & 512(447) & 698(449)  &  8e-31\\   
1-1.2&  $0.36\pm0.01$ & $-$ & $\leq257$ & $2.4\pm0.3$ & $0.47\pm0.03$ & $1.0\pm0.1$ & 433(415) & 623(417) & 2e-33\\   
7-8& $0.41\pm0.03$ & $-$ & $\leq257$ & $1.8\pm0.2$ & $0.32\pm0.03$  & $0.3\pm0.1$ & 382(289)  & 422(291) & 6e-7 \\ 
\hline
\end{tabular}
\end{table*}

\subsection{The prompt SED}
\label{s:promptsed}
The quasi-simultaneous optical and gamma-rays flaring activity at $\sim T_0+2$ ks (\S \ref{s:lag}) may suggest a common origin of the emission mechanism responsible for the optical and high energy radiation. We further investigate this hypothesis by analyzing the simultaneous optical to X-ray Spectral Energy Distribution (SED). 
To this aim we extract the {\it Swift}/UVOT, XRT and BAT data energy spectrum in three temporal intervals. We selected the ranges 650-850 s and 1-1.2 ks after $T_0$ since they were covered in a large frequency window and free of flaring activity (Fig.\ref{figure:all}). In addition, we extract a third SED at 7-8 ks after $T_0$ where {\it Swift}/UVOT and XRT energy spectra only were available. 

First, we simultaneously fit the XRT and BAT data assuming a broken power law and a Band spectral model. 
As clearly shown in Figure \ref{figure:promptsed} for the broken power law case, while the 15-150 keV spectrum is consistent with the  extrapolation from the 0.3-10 keV spectrum to higher energies, the simultaneous TAROT and {\it Swift}/UVOT R, {\it v, b, u, w1, m2} and {\it w2} SED provides a measured softer spectrum, with spectral index $\beta_{opt}=1.43\pm0.20$, at odds with the expectations where both X-rays and optical emission is synchrotron radiation from the same population of electrons with a power law spectral energy distribution. Indeed, for synchrotron models, the spectral slope always gets steeper (thus softer) with increasing frequency. 

In analogy with the study of the prompt emission spectral continuum of other LGRBs (e.g. GRB 100901A, Gorbovskoy et al. 2012), a non negligible absorption by dust in the host galaxy allows to recover the optical with the X-ray spectrum. We test this hypothesis by simultaneously fitting BAT, XRT, UVOT and TAROT data\footnote{We used the \texttt{flx2xsp} tool to build the optical flux spectrum. } assuming three possible dust extinction laws (the Galactic and the two Magellanic Clouds ones) using the software XSPEC \citep{Arnaud1996} v12.8.0, and we find acceptable solutions (Tab. \ref{table:prompt}). Assuming a Band model rather than a broken power law model does not improve significantly the goodness of the fit. In both cases (Band and broken power law model), we find that the Small Magellanic Cloud dust extinction curve best fit the data, against the Milky Way (MW) and the Large Magellanic Cloud (LMC) laws. In addition, for both spectral continuum models, the fit improves with the addition of a black body component in the soft X - far UV energy range, 
consistent with the spectral analysis results obtained from fitting the XRT data only published in Paper I. In Table \ref{table:prompt} we quote the statistical significance of the fit improvements with the addition of a black body spectral component.

From the KW data spectrum integrated over the main burst, that is from $T_0-1990$ to $T_0+4400$ s, we learn that the best fit spectral model in the 20-1400 keV energy range is a cut-off power law with spectral index $\beta_{KW}=0.31\pm0.09$ (photon index $\Gamma_{KW}=1.31$) and cut-off energy $E_{cut}=310\pm53$ keV \citep{Golenetskii2011}. This model is better in agreement with the well known $E_{iso}-E_{peak}$ correlation \citep{Amati2002,Amati2010} according to which the expected $\nu F\nu$ peak energy value is about few hundreds of keV given the large equivalent isotropic energy value for this burst of $E_{iso}=(5.7\pm0.7)\times10^{53}$ erg \citep{Golenetskii2011}. Therefore, by assuming that the bulk of the radiation for this burst is emitted at energies above the BAT energy range (i.e. $E_{peak}>150(1+z)$ keV), we test the KW exponential cut-off power law model on the simultaneous {\it Swift}/XRT and BAT spectrum by fixing the photon index and cut-off energy to vary within their uncertainties, that is with a cut-off energy in the range 257-363 keV and a spectral index in the range 0.22-0.44, and letting free to vary the normalization. The best fit model can reproduce our estimate of the KW flux density at $2.8\times10^{19}$ Hz (i.e. 116 keV, see \S \ref{s:datared}), that is $\sim9\times10^{-5}$ Jy at $T-T_0=650-850$ s and $\sim4\times10^{-5}$ Jy at $T-T_0=1-1.2$ ks, but could only marginally fit the XRT and BAT data  ($\chi^2/$d.o.f.$=676/443$). By introducing a black body component we find a significant improvement of the fit ($\chi^2/$d.o.f.$=500/441$) with $kT\sim 1$ keV. By assuming two black body components, we find further improvement to the fit with the second thermal component in the FUV energy range ($\chi^2/$d.o.f.$=466/439$). 

Also in this case, however, the optical fluxes are severely underpredicted by this model (Fig. \ref{figure:promptsed}).  Given the harder photon index inferred from the cut-off power law model, the extrapolation to the optical range requires in this case a larger amount of dust extinction. In addition, a different dust extinction law, more similar to the MW one rather than the SMC, seems to better represent the observations. 

Therefore, independendently on the assumed intrinsic spectral model, we find that the optical emission is consistent with the X-ray/gamma-ray spectral continuum only if a non-negligible amount of dust is introduced. 

\subsection{Dust to gas ratio}

A considerable amount of dust along the line of sight should reflect in the X-ray data analysis with a certain amount of equivalent hydrogen (gas) column. 
Past studies have shown that on average the $N_H/A_V$ measured ratio in optically bright GRBs is about 10 higher than the expected one \citep[e.g.][]{Galama2001,Stratta2004,Kann2006,Schady2010}, although with a large scatter. 

The average value of $<N_H/A_V>=(3.3\pm2.8)\times10^{22}$ has been measured for a sample of optically bright GRBs assuming a SMC environment \citep{Schady2010}. From the measured $N_H$ obtained during prompt emission analysis (Tab.\ref{table:prompt}) we can estimate an expected range of dust extinction values of $A_{V,exp}=N_H/<N_H/A_V>=[0.04-0.8]$ mag. These values well includes the measured $A_V=R_V\times E(B-V)$ that we find in the range $\sim0.3-0.85$ mag assuming a broken power law and a Band model. For these models we find a best fit for an SMC dust extinction law that requires a total-to-selective extinction parameter $R_V$ of $2.93$ \citep{Pei1992}. 
 
By assuming a cut-off power law spectral model for the prompt spectral continuum, the extrapolation up to the optical range requires an MW-like extinction law with $A_V$ values estimated in the range of 0.9-1.5 mag for $R_V=3.1$ \citep{Pei1992}. Assuming an MW extinction curve, the typical GRB average $<N_H/A_V>$ is $(2.1\pm1.8)\times10^{22}$ (Schady et al. 2010). From the measured $N_H$ values (Tab.\ref{table:prompt}), we  expect $A_V$ of 0.05-1 mag, that is consistent with the lowest values of our estimated $A_V$ range. More consistent values are obtained if we assume the empirical $N_H/A_V$ ratio measured in our Galaxy, that is $<N_H/A_V>_{MW}=1.8\times 10^{21}$ \citep{Predehl1995}.   

Given the better consistency of a cut-off power law model both with the KW data analysis \citep{Golenetskii2011} and with the expected $E_{iso}-E_{peak}$ correlation for LGRBs, we tentatively conclude that a dust to gas ratio similar to that one observed in our Galaxy is preferred for GRB 111209A.


   \begin{figure*}
   \centering
   \includegraphics[width=9cm]{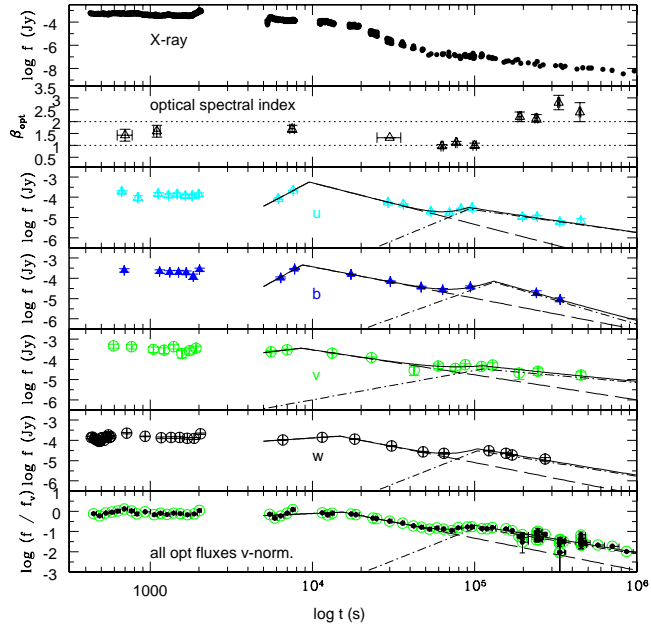}
   \caption{{\it Swift}/UVOT {\it white, u, b} and {\it v} filter light curves as well as the {\it Swift}/UVOT co-added light curve normalized to the {\it v}-band (bottom panel), compared with the {\it Swift}/XRT light curve (top panel) and the optical spectral indices versus time from trigger. All the light curves have been fitted in the temporal range 3-1000 ks from the BAT trigger by the sum (continuous line) of two broken power laws (long dashed line and dot-dashed line).  }
              \label{f:otlc}%
    \end{figure*}
%

\begin{table*}
        \centering
                \begin{tabular}{l c c c c c c }
\hline
band & $\alpha_1$ & $t_{b}$ & $\alpha_2$ & $\alpha_{r,1}$ & $t_{b,r}$ & $\alpha_{r,2}$ \\
        &                       & ks &                  &                                       & ks    &                       \\
\hline
{\it u}       & $\sim-4.2$ & $\sim10$                 & $\sim2.0$             & $\sim-3.3$ & $92\pm5$  & $1.1\pm0.2$ \\
{\it b}       & $\sim-4.4$ & $\sim9$          & $\sim1.5$             & $\sim-2.9$  & $\sim131$& $2.3\pm0.3$  \\
{\it v}       & $\sim-0.9$ & $8\pm1$          & $1.2\pm0.2$   & $\sim-1.4$ & $113\pm10$& $0.7\pm0.4$ \\ 
{\it w}       & $-0.5\pm0.1$& $14.8\pm0.5$ & $1.6\pm0.1$   & $\sim-3.1$ & $104$     & $1.3^{+0.2}_{-0.6}$ \\
mean & $-0.5\pm0.1$ & $13.4\pm0.4$ & $1.5\pm0.1$  & $-2.0\pm0.5$ & $96\pm4$ & $1.3\pm0.1$ \\
\hline
{\it v}-norm & $-0.35\pm0.10$ & $15.5\pm0.4$ & $1.6\pm0.1$  & $-2.0\pm0.5$ & $102\pm2$ & $1.3\pm0.1$ \\
\hline
\end{tabular}
\caption{{\it Swift}/UVOT best fit parameters of the light curves in the temporal interval 3-1000 ks by assuming two broken power laws (where each power law segment is represented by $F(t)\propto t^{-\alpha}$). The ``on-set bump" rising and decay indices, as well as its peak epoch are $\alpha_1$, $\alpha_2$, and $t_b$. The suffix ``r" indicates the same parameters for the late rebrightening feature. The row ``mean" quotes the weighted mean of each parameter measured in the optical range, where for those values for which we could not compute the uncertainties due to poor statistics we considered a relative uncertainty of $50\%$. The row ``{\it v}-norm" quotes the results obtained from the light curve obtained with all filters, normalized to the v-band fluxes (see text). }
\label{tab:otlc}
\end{table*}

\section{The afterglow emission}
\label{s:afterglow}

In this section we analyze the prompt to afterglow transition and the afterglow properties through a multiwavelength data analysis using {\it Swift}/UVOT and XRT data, high quality data from XMM-Newton and ATCA radio fluxes.  In the following, we will assume the standard paradigm in which the steep decay phase observed in X-rays to start at about $T_0+20$ ks is the evidence of high latitude emission from the prompt phase \citep{Kumar2000}, while the following shallow decay observed in the XMM-Newton data, and the normal decay observed also in the {\it Swift}/XRT data, represent the afterglow emission.

   \begin{figure*}
   \centering
   \includegraphics[width=8cm,angle=0]{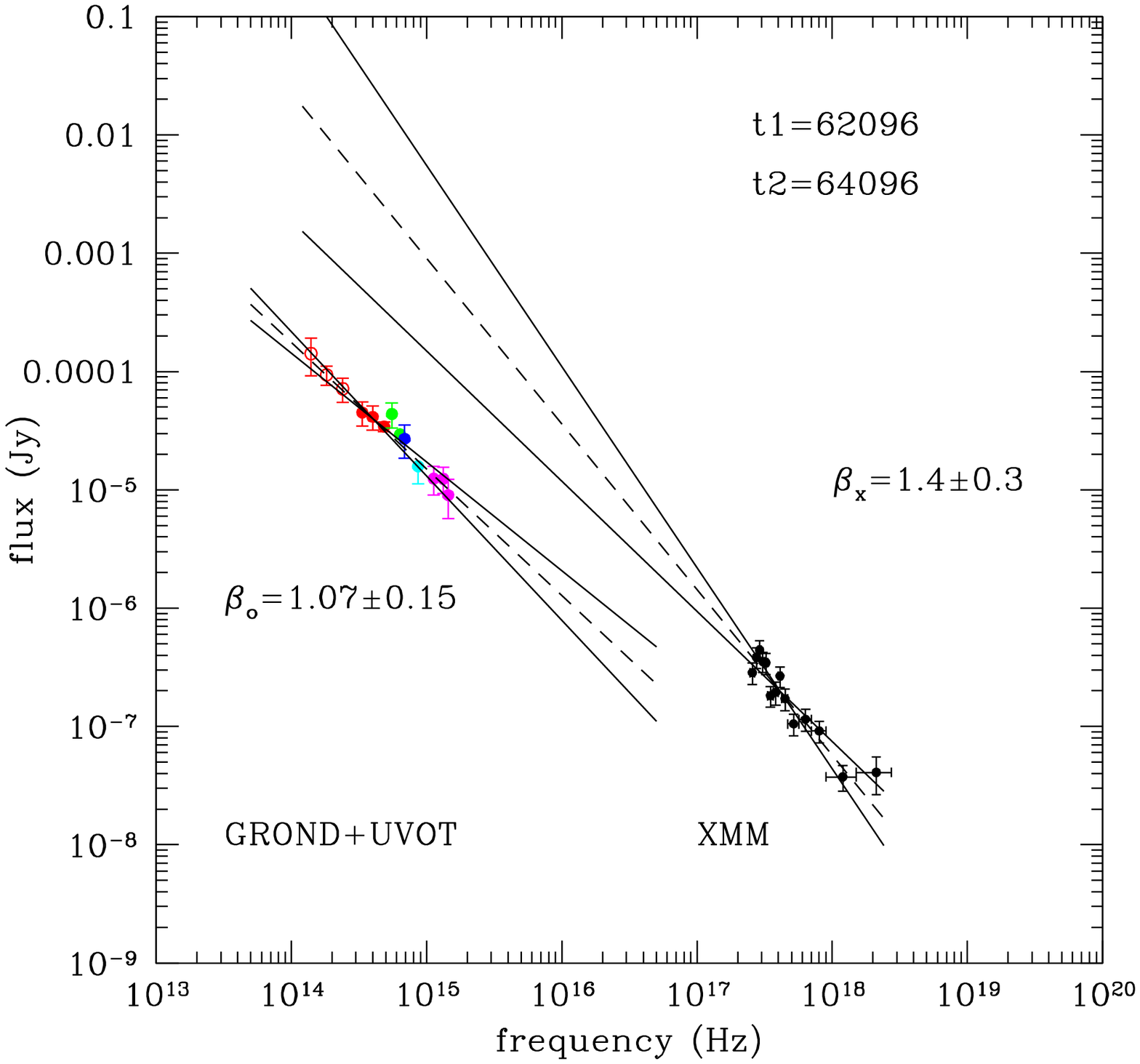}   
   \caption{The Spectral Energy Distribution at the GROND data epoch at $T_0+63$ ks (17 hours after the trigger), where the grizJHK GROND data are plotted together with the simultaneous {\it Swift}/UVOT and XMM data.}
              \label{figure:sedgrond}%
    \end{figure*}

\subsection{{\it Swift}/UVOT optical light curve and spectral analysis}
\label{s:opt}
 
Starting from the second {\it Swift} orbit after the trigger, that is from $\sim T_0+3$ ks, the optical flux shows a marginal increase at all wavelengths up to about 10 ks after the trigger. This may suggest that a different emission component is emerging over the end part of the prompt emission. This phase is then followed by a smooth power law decay up to about 1 day after $T_0$. Another rebrightening is then observed to peak around $T_0+1.16$  days (e.g. Fig.\ref{figure:all}, see also  Kann \& Greiner 2011b). 

In order to better quantify this behavior, we fit the light curves extracted in each UVOT filter assuming a model formed by the sum of two broken power laws, in the temporal range that goes from $T_0+3$ ks to $T_0+1$ Ms ($\sim 10$ days).  
To have a prompt comparison with the X-ray light curve, we plot the latter in the same figure. 
Not all the model parameters could be quantified by fitting the light curve extracted in each separate UVOT filter. Therefore, to increase the available statistics and better define the model parameters, we normalize all fluxes to the {\it v}-band flux following \cite{Oates2009} using data between 20 ks and 70 ks, under the assumption that spectral variability is not significant across the light curve. This assumption is supported by the fact that no spectral  dependence of $\alpha_2$ is detected, see Table \ref{tab:otlc}. We then grouped the resulting light curve with a time resolution of $\delta t/t \sim0.1$. Results for the {\it u, b, v}, as well as for the {\it white} filters and for the coadded and {\it v}-band normalized light curves are quoted in Table  \ref{tab:otlc} and plotted in Figure \ref{f:otlc} together with the X-ray light curve (top panel). 

From the {\it v}-band normalized optical light curve the rising flux behavior around 5-10 ks is very steep, with a power law index of $-3.7\pm0.3$, and it is preceded by a marginal evidence of a rapid flux decay with index $\sim 1$. Considering that at this time the gamma-ray emission is still active, this result may indicate a possible contamination from a non negligible optical flaring activity, preventing a careful analysis of the optical rising phase and of the peak epoch estimate. 
On the contrary, after $\sim T_0+10$ ks the decay slope shows a more consistent value at all wavelengths within the uncertainties. Results are quoted in Table \ref{tab:otlc}.

We then extract the spectral information at several epochs. The latter ones were selected in order to minimize the uncertainties introduced by the necessary temporal extrapolation at each specific epoch of the sparse data sample.  We apply a smooth polynomial function to approximate the light curve in each band and then we compute the average value for each selected temporal interval and for each filter. Between 7 and 8 ks, and then between 20 and 40 ks after $T_0$, we measure an energy spectral index of $\beta_{opt}= 1.60\pm0.25$ and $\beta_{opt}= 1.33\pm0.01$, respectively using {\it Swift}/UVOT data.  
We note that at a redshift z=0.677 the UVOT frequencies trace the rest frame UV emitted radiation and, given the large sensibility of this energy range to even small amount of dust along the line of sight, the true spectral slope may be harder than the measured value. 

On the contrary, just before the re-brightening we could measure the spectral index using also NIR data taken with the GROND telescope at $T_0+63$ ks (Fig. \ref{figure:sedgrond}). The resulting optical/NIR SED is well fitted by a simple power law with spectral index of 
$\beta_{opt}=1.07\pm0.15$. 
Another epoch where we could exploit simultaneous NIR data is around 77 ks where  VLT g, R, i and z data are available in addition to the {\it Swift}/UVOT, and we measure $\beta_{opt}=1.11\pm0.06$. 

Around the re-brightening peak, between $90$ and 100 ks after $T_0$, the optical/UV spectral index is $\beta_{opt}=1.0\pm0.1$. After the peak, the optical/UV spectral slopes soften significantly. We could measure the optical spectral index at 4 post-rebrightening epochs where {\it Swift}/UVOT data were taken nearly simultaneous in all filters. We find $\beta_{opt}=1.2\pm0.2$ at $T_0+190$ ks, $\beta_{opt}=1.14\pm0.16$ at $T_0+240$ ks, $\beta_{opt}=1.8\pm0.3$ at $T_0+330$ ks and  $\beta_{opt}=1.4\pm0.4$ at $T_0+446$ ks that is at the radio epoch data (see \S \ref{s:radiosed}). 

An early ``bump" and a late re-brightening have been observed in several ``normally" long GRBs. \cite{Liang2012} provide an extensive data analysis of these optical features for a large sample of 146 optically monitored long GRBs. In 38 and 27 GRBs, on-set bumps and re-brightenings have been detected, respectively. In 12 cases both features were present in each optical light curve. The rising index distribution of the on-set features ranges from -0.3 to -4 with a mean at about -1.5 and it is consistent with the rising index distribution of the late rebrightening.  The decay indices distribution of both the on-set bumps and late rebrightenings ranges whithin 0.6-1.8 with a mean at 1.15, apart from three outliers with decay index grater than 2. The on-set optical bump peak epoch usually coincides with the X-ray light curve entering in the steep decay phase, suggesting a different origin  of the radiation observed in the two energy domains at that time, while late optical rebrightenings are on average tracked also in X-rays. 

We find that the optical afterglow of GRB 111209A fits with the average behavior, 
the only marked difference is in the peak time of the on-set bump. Indeed, while the late re-brightening peak epoch is consistent with the poorly defined typical range that goes from several hundreds of seconds to days after the burst trigger, the on-set bump of GRB 111209A peaks at much later times than the typical 30-3000s on-set peak epoch after the burst trigger time. These findings support our impression that the prompt emission contamination is indeed preventing us the correct estimate of the rising index and the peak epoch of the early optical emission. We discuss some possible interpretations of the optical afterglow feature origin in \S \ref{s:sed}


   \begin{figure*}
   \centering
   \includegraphics[width=8cm,angle=0]{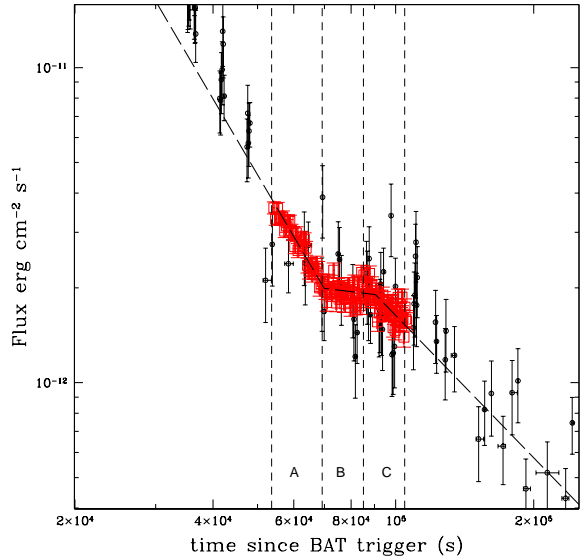}
      \caption{{\it Swift}/XRT PC (black open circles) and the XMM-Newton PN (red open squares) X-ray flux light curve. The dashed line is the best fit obtained from simultaneous XMM-Newton and {\it Swift}/XRT PC data.}
         \label{f:xlc}
   \end{figure*}

   \begin{figure}
   \centering
   \includegraphics[width=10cm,angle=0]{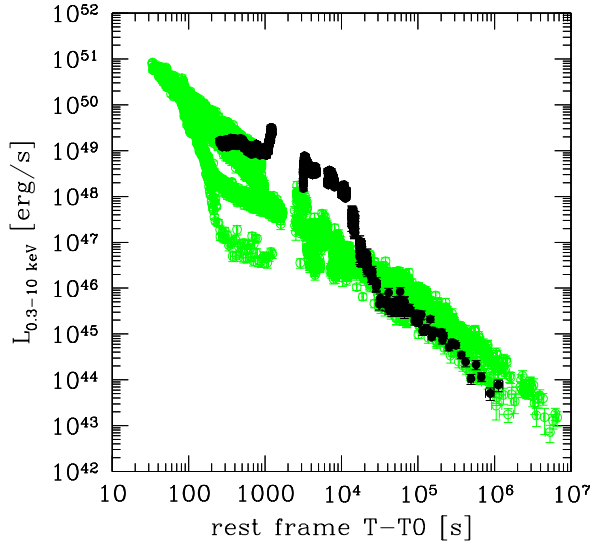}   
   \caption{X-ray (0.3-10 keV) rest frame luminosity of GRB 111209A (black) and of a sample of LGRBs at known  redshift and consistent with the ICG model (green) taken from \cite{Pisani2013}.}
              \label{f:igc}%
    \end{figure}

   \begin{figure*}
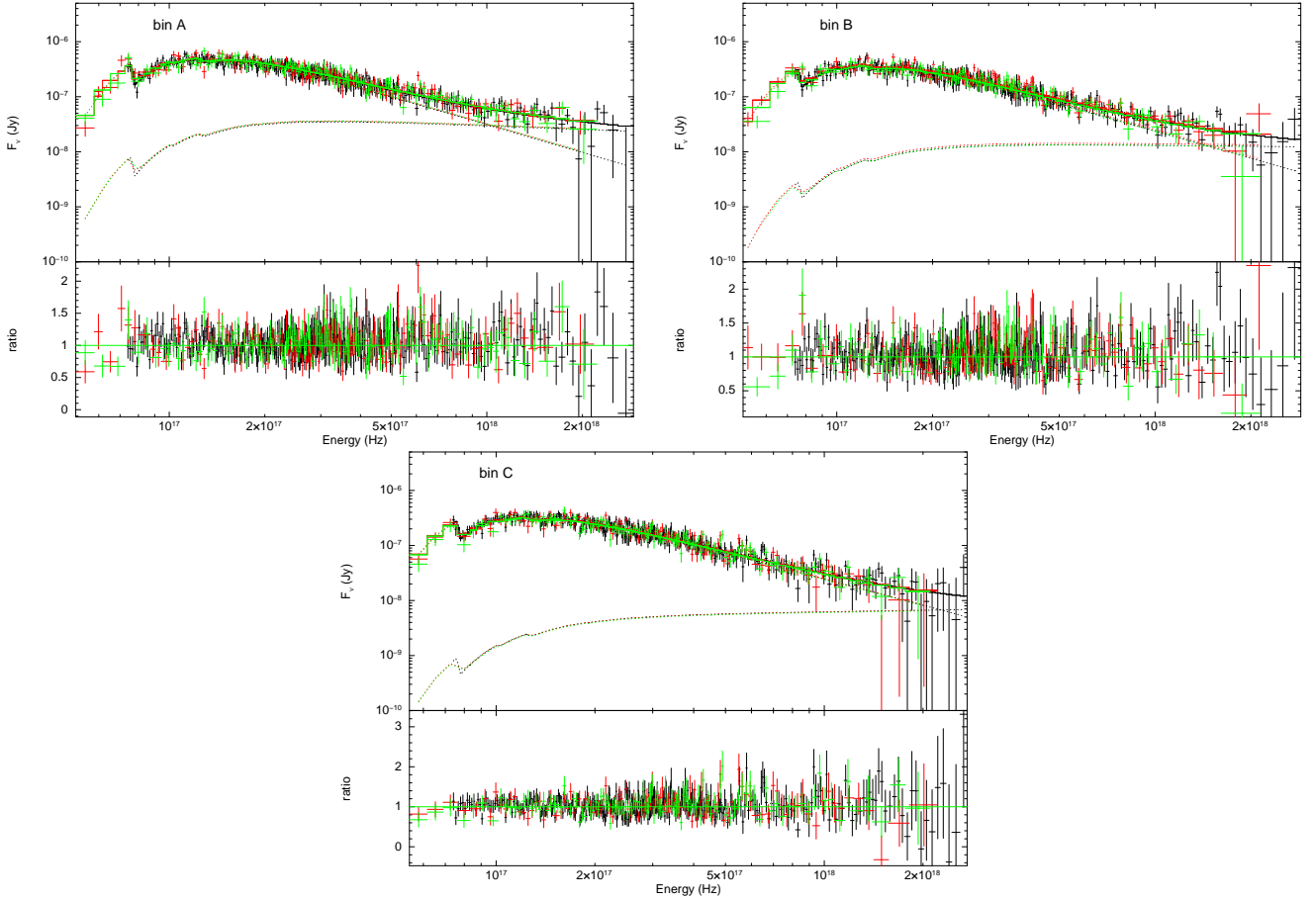

   \centering
\includegraphics[width=6cm,angle=-90]{fig8a_new.eps}
\includegraphics[width=6cm,angle=-90]{fig8b_new.eps}
\includegraphics[width=6cm,angle=-90]{fig8c_new.eps}
      \caption{XMM PN and MOS 1 and 2 spectra extracted at three different epochs (during the steep phase, the plateau phase and after the end of the plateau phase), fitted with 2 absorbed power laws. The three temporal bins are marked as ``A", ``B" and ``C" in Figure \ref{f:xlc}. Best fit parameters are quoted in Table \ref{table:xmm}.}
         \label{f:xmm}
   \end{figure*}

\begin{table*}
\caption{XMM data spectral analysis}             
\label{table:xmm}      
\centering                          
\begin{tabular}{c c c c c c c c c c}        
\hline\hline                 
bin & Time range & Model & $\beta_1$ & $\beta_2$ & $N_{H,z}$ & $\chi^2$(d.o.f.) & $F_1(F_{1,unabs})$ & $F_2(F_{2,unabs})$ & F-test \\    
    &  ks           &            &               &       & $10^{21}$cm$^{-2}$ & & \multicolumn{2}{c}{(0.3-12 keV) $10^{-12}$ cgs}  &  \\
\hline                        
A & 56.7-70.0 & PL & $1.30\pm0.05$ & - &$1.6\pm0.2$ & 650(604) & 2.2(2.7) & - & -\\     
B &  70.0-87.7 & PL & $1.40\pm0.06$ & - &$1.5\pm0.2$ & 583(537) & 1.4(1.9) & - & -\\    
C &  87.7-108.2 & PL & $1.4\pm0.1$ & - &$1.8\pm0.2$ & 600(586) & 1.3(1.8) & - & -\\    
\hline                                   
A & 56.7-70.0 &2PL& $1.7^{+0.8}_{-0.3}$ & $0.3^{+0.6}_{-1.0}$ & $2.4^{+0.4}_{-0.7}$ & 615(602)& 1.5(2.5) & 0.9(0.9) & 6e-8\\
B & 70.0-87.7 &2PL& $1.6^{+0.7}_{-0.2}$   & $0.0^{+2.3}_{-1.3}$ & $2.0^{+0.6}_{-0.3}$ & 565(535)& 1.2(1.8) & 0.3(0.3)& 2e-4 \\   
C & 87.7-108.2 &2PL& $1.6^{+2.0}_{-0.1}$& $-0.1^{+3.7}_{-3.1}$ & $2.0^{+3.2}_{-0.3}$ & 595(584)& 1.2(1.8) & 0.2(0.2)& 0.1 \\   
\hline                                   
\end{tabular}
\end{table*}

\subsection{The late afterglow as observed by XMM-Newton}
\label{s:xmm}

The exceptionally high quality of XMM data enabled us to detect the presence of a plateau at the end of the steep decay with more statistical confidence than using Swift/XRT data alone. Assuming a broken power law model, with an initial steep decay index of $\alpha_{1,XMM}=2.23\pm0.10$, a break at $t_{b,1}=70\pm1$ ks followed by a shallower decay with index $\alpha_{2,XMM}=0.53\pm0.05$, could only marginally reproduce the data ($\chi^2=143/100$). By extrapolating this model at late times, it clearly overpredicts the ending part of the XMM-Newton data as well as the following {\it Swift}/XRT PC fluxes. A significant $\chi^2$ improvement was obtained by allowing the presence of a second temporal break at $t_{b,2}=87\pm1$ ks ($\chi^2=124/98$).   
Performing an F-test to evaluate the statistical significance of the addition of a second temporal break, we find a null-hypothesis probability value (P-value) of 0.001. A simultaneous fit of the XMM-Newton data and the {\it Swift}/XRT PC data from $T_0+100$ ks up to the {\it Swift}/XRT monitoring end ($T_0+2$ Ms), provides a better constraint of the final decay slope, with $\alpha_{3,XMM+XRT}=1.52\pm0.06$, and a shallow decay phase with $\alpha_{2,XMM+XRT}=0.18\pm0.05$  (Fig.\ref{f:xlc}). 

Shallow decay phases, or plateaus, are commonly observed in GRB X-ray light curves with {\it Swift}/XRT \citep{Liang2010}. A characteristic property of X-ray plateaus is that the X-ray luminosity at the end of the plateau typically scales with the rest frame epoch of the plateau end as  $L_X\sim10^{51}T_{end}^{-1}$ erg s$^{-1}$,  although with a large scatter \citep{Dainotti2010}. For GRB 111209A, we measure $L_X\sim4\times10^{45}$ erg s$^{-1}$ at the rest frame epoch of the plateau $T_{end}=t_{b,2}/(1+z)\sim52$ ks after $T_0$ (Fig.\ref{f:igc}). These values make GRB 111209A marginally consistent with the correlation within its large intrinsic scatter, and put it at the bottom-right end of the $L_X-T_{end}$ plane where are those GRBs with the faintest and longest plateaus.

We divided the XMM observations into three temporal bins tracking the ``steep-flat-steep" phases, labelled as  ``A", ``B" and ``C" in Figure \ref{f:xlc}. We then fitted the extracted X-ray spectra with a simple and a double power law model. Results are quoted in Table \ref{table:xmm}. Uncertainties are at $90\%$ confidence level. 
The resulting 0.3-12.0 keV PN energy spectrum and the two 0.2-10.0 keV MOS spectra, all grouped in order to have at least 20 counts per energy bin and fitted simultaneously, are plotted in Figure \ref{f:xmm}.  
During the first (``A") and only marginally on the second (``B") temporal bin, data are better fitted by the addition of a second harder component to the standard power law soft spectrum usually detected during this phase with {\it Swift}/XRT data. Performing an F-test, the statistically significance of the additional hard power law is measured with a null hypothesis probability of $10^{-8}$ and $10^{-4}$ during bins ``A" and ``B", respectively (Tab. \ref{table:xmm}). We discuss on the possible origins of this component in \S \ref{s:discussion}.


   \begin{figure*}
   \centering       
 \includegraphics[width=8cm,angle=0]{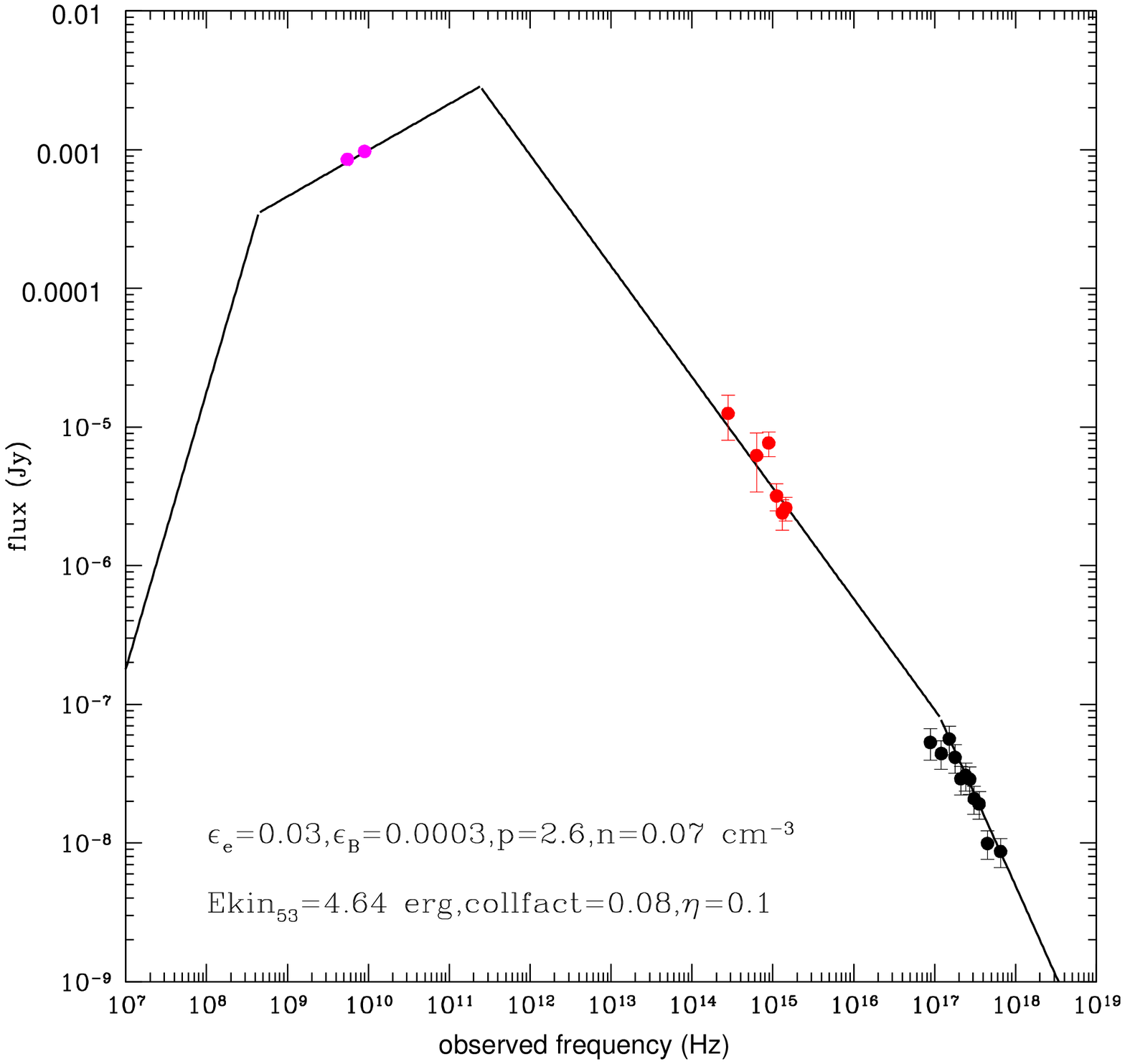}           
   \caption{The simultaneous radio (magenta filled points), {\it Swift}/UVOT m2,w1,u,v and v band, and {\it Swift}/XRT X-ray data Spectral Energy Distribution at the epoch of radio observations (around 5 days after the trigger). The solid line is the solution find in the context of the synchrotron emission model following  \cite{Panaitescu2000}.}
              \label{figure:sedradio}%
    \end{figure*}

\subsection{The late radio to X-ray afterglow SED}
\label{s:radiosed}

Nearly simultaneous radio, X-rays and optical data in the {\it w2, m2, w1, u, b} and {\it v} filters are available at $T_0+5.2 $ days. We corrected the optical flux from Galactic absorption and we plot them together with the unabsorbed X-ray and radio data (Fig.\ref{figure:sedradio}). 
The X-ray spectrum was extracted in the temporal range that goes from $T_0+250$ ks to $T_0+650$ ks using the {\it Swift}/XRT spectra repository time-sliced spectral analysis tool. The X-ray spectral index, $\beta_X=1.8\pm0.4$ and decay index of $\alpha_X=1.5\pm0.1$, are consistent with the synchrotron model expectations for $\nu_X>\nu_c$, where $\nu_c$ is the cooling frequency of the electrons \citep{Sari1998}.  
By fixing the optical spectral index free to vary within $\beta_X$ and $\beta_X-0.5$ (including the uncertainties on $\beta_X$), the optical to X-ray data are best fit by a broken power law model with $\beta_X=1.6\pm0.4$, $\beta_{opt}=0.87\pm0.03$, a break energy in the range $0.6-1.4$ keV and no rest frame visual dust extinction ($\chi^2=10.5$ for 10 degrees of freedom). Despite the large uncertainty affecting the X-ray spectral slope, a simple power law spectral continuum between the X-rays and the optical frequencies provides a much worse fit ($\chi^2=19$ for 12 degrees of freedom) and we can confidently exclude this model. 
Radio data alone provide excellent agreement with the expected 1/3 spectral slope if $\nu_{radio}$ were below the spectral peak frequency. Following \cite{Panaitescu2000}, we could fit the multi-band SED  
for $\nu_{radio}<\nu_m<\nu_{opt}<\nu_c<\nu_X$, where $\nu_m$ is the synchrotron injection frequency \citep{Sari1998}, with an environment density $n=0.07$ cm$^{-3}$, a fraction of the total energy tranferred to the magnetic field of $\epsilon_B=0.0003$ and to the swept-up electrons of $\epsilon_e=0.03$, $p=2.6$, $\eta=0.1$ and a collimation factor of 0.08 (i.e. a half opening angle of 23 degrees). We could not find any obvious solution assuming a wind environment.

\section{Discussion}
\label{s:discussion}

\subsection{Evidence of dust destruction?}
\label{s:dust}

We have seen in \S\ref{s:promptsed} that to interpret the prompt optical emission observed by {\it Swift}/UVOT and TAROT as originating from the same physical mechanism responsible of the observed emission in X-rays and gamma-rays, we are forced to invoke a non negligible amount of dust.  

While invoking dust extinction is not rare in the afterglow spectral analysis, given our ignorance on the physical radiative mechanism at the basis of the prompt emission, the possibility that a distinct phenomenology is mimicked by dust extinction cannot be excluded. In particular, for GRB 111209A we have seen that at about $T_0+2$ ks a flare clearly detected with TAROT and Konus-Wind shows an optical peak epoch delayed of about 400 s from the gamma-ray peak. If all the optical emission is delayed with respect to the high energy counterpart, the simultaneous prompt emission SED should take into account such temporal lag. We test this scenario by simultaneously fitting X-ray spectrum extracted at $T_0+700$ s with optical spectrum at $T_0+1100$ s, but even in this case, we find that the optical fluxes are severely underestimated and a non-negligible dust extinction may recover the expected values. 
Another way to mimick a dust extinction effect is the presence of a spectral break between the X-rays and the optical energy domains. However, we have already considered this case in our heuristic broken power law spectral model, with a break energy in or below the soft X-ray energy range, and we have shown that the much softer optical spectral index can recover the expected harder optical to X-ray spectral slope with the presence of dust extinction. 

Interestingly, non negligible dust extinction during the prompt emission has already been invoked for another very long GRB whose duration falls near the right end of the burst duration distribution, that is, GRB 100901A with $T_{90}=439\pm33$ s \citep{Sakamoto2010}. In that case, to match the optical data with the X-ray extrapolation during the prompt emission where simultaneous optical to gamma-ray spectrum were available, a rest frame visual dust extinction of $A_V\sim0.6-1.6$ was introduced \citep{Gorbovskoy2012}. In addition, two dark GRBs with $A_V>5$ mag in the rest frame were found to have a very long duration prompt emission, with $T_{90}\sim 800$ s, although other long GRBs with $T_{90}>500$ s were found with no evidence of strong dust extinction \citep{Zauderer2013}. 

However, for GRB 111209A, we find that gathering evidence of dust extinction from the afterglow emission is more contrived. 
The best measure we could perform of any dust reddening in the afterglow emission was obtained at $T_0+63$ ks when simultaneous GROND and {\it Swift}/UVOT data are available and provide a NIR-to-UV SED. We could not find any evidence of large dust extinction effects since they are well fit by a simple power law model ($\chi^2=9.5$ with 10 degrees of freedom), with a $90\%$ upper limit of $A_V<0.3$ mag. In Figure \ref{figure:sedgrond} we plot the simultaneous NIR/optical and X-ray spectra. Despite the spectral slopes in the two energy regimes are consistent within their uncertainties, there is a clear normalization mismatch. This is not surprising since at this epoch the X-ray emission is likely still affected by the prompt emission\footnote{Note that the X-ray spectrum plotted in Figure \ref{figure:sedgrond} is integrated over the small temporal interval $t_2-t_1$ about the GROND observation epoch and collected photons provide insufficient statistics to detect the hard component (see \S \ref{s:xmm}).}. Even neglecting any prompt contribution, we find that the two spectra can be modelled with a broken power law ($\chi^2/d.o.f.=19/15$), with a 2$\sigma$ upper limit of $A_V<0.15$. 
At $T_0+77$ ks a simultaneous SEDs using XMM-Newton and UVOT data as well as VLT g, r, i and z observations from Levan et al. (2013) were available and again we find similar results, with a rest frame dust extinction of  $A_V<0.1$ mag.

From the gas column measured using XMM-Newton data in the rest frame of the burst, we find a range of values $1-4\times10^{21}$ cm$^{-2}$ (Tab.\ref{table:xmm}) by assuming a solar abundance and a cold environment. These values provide an expected dust extincion $A_{V,exp}$ of the order of $\leq0.1$ mag assuming the average GRB $<N_H/A_V>=(2.1\pm1.8)\times10^{22}$ for a MW environment (Schady et al. 2010) or $A_{V,exp}=[0.6-2.2]$ mag assuming the empirical $N_H/A_V$ observed in our Galaxy. Finally, from the radio to X-ray SED extracted at 443 ks after $T_0$, we find no evidence of dust extinction.

The low dust extinction level inferred from the optical afterglow analysis and the contrasts with the prompt emission findings and with the expected $A_V$ from the XMM-Newton data analysis, can be reconciliated by assuming an effective dust destruction mechanism at play during the prompt and the afterglow emission. In this case, the dust inferred during the prompt emission should be located not too distant from the central engine. Dust destruction simulations show that intense GRB fluxes can destroy dust up to a radius of $\sim 10$ pc \citep[e.g.][]{Waxman2000}, consistent with a dense and dusty star forming region in which the GRB is embedded, although the possibility that the dust is produced by the progenitor star itself can not be excluded. Dust destruction is a likely possibility given the intense UV and soft X-ray photon fluxes from a GRB  and possible evidence of such mechanism has been recently suggested for GRB 120119 during the final phases of the prompt emission \citep{Morgan2013}, exactly as we propose to be the case for GRB 111209A.

\subsection{Implications for the BSG progenitor}
\label{s:progenitor}
The presence of dust in the host galaxy of GRB 111209A is in line with the recent findings of a subsolar, but not especially low host galaxy metallicity. From $12+log(O/H)=8.3\pm0.3$ (Levan et al. (2013) and assuming $log(Z/Z_{odot})=12+log(O/H)-8.76$ \citep{Caffau2008}, we inferr $\sim0.35 Z_{\odot}$. At the same time, since for short lived sources such as massive stars the host metallicity likely reflects the star metallicity, 
this result is at odds with a BSG progenitor interpretation \citep{Gendre2013,Nakauchi2013}. Indeed, following \cite{Woosley2012}, a low metallicity condition of $<0.1 Z_{\odot}$  was invoked for this progenitor star (Paper I) in order to prevent strong wind and thus to provide sufficient mass to supply the central engine over several hours. 

A possible solution can be found in a binary system formation channel of the blue supergiant \citep[e.g.][]{Podsiadlowski1992}. Binary systems, either formed by two massive stars or a massive star and a low mass companion   \citep{Fryer2005,Podsiadlowski2010}, have been invoked to deal with the increasing evidence that a large fraction of LGRBs explode in high metallicity environments, that is, where the required high core angular momentum condition is expected to be suppressed by strong winds \citep [see a recent review by e.g.][and reference therein]{Levesque2013}. 
A binary system composed by a Helium star and a neutron star (NS) within a common envelope phase \citep{Fryer1998}, has been suggested for GRB 101225A, another ultra-long burst (with bursts duration $>2000$ s), that share with GRB 111209A similar host galaxy properties \citep{Levan2013}, but for which a clear thermal component was detected in the optical afterglow between 1 and several tens of days after the trigger \citep{Thoene2011}, contrary to what we see for GRB 111209A. 

Another interesting scenario suggested by a binary system progenitor for LGRBs, is the ``Induced Gravitational Collapse" (IGC) of a NS into a black hole (BH) through accretion from a massive companion during its supernova (SN) phase  \citep{Ruffini2001}. This scenario could explain a number of LGRBs with burst durations ($T_{90}$) in the range $\sim10-100$ s, and their afterglow phenomenology \citep{Izzo2012,Pennacchioni2012,Pennacchioni2013}. The IGC  model predicts a number of features: a precursor in the prompt emission light curve due to the initial phase of the SN, an optical SN Ic signature tens of days after the burst onset and a universal rest frame 0.3-10 keV light curve at late times \citep{Pisani2013}, possibly due to the newly-born NS after the SN phase of the donor star.  Whether this model applies to the ultra-long GRB 111209A requires accurate modelling. However, the Konus-Wind light curve of GRB 111209A clearly shows a precursor\footnote{http://www.ioffe.rssi.ru/LEA/GRBs/GRB111209A} at about $\sim T_0-10$ ks. The detection of a SN for GRB 111209A is not obvious from the late time optical light curve, although possible indirect evidence has been discussed by Levan et al. (2013). Finally, by comparing the sample of IGC-GRBs at known redshift \citep{Pisani2013} with GRB 111209A, we find that the late X-ray afterglow luminosity of GRB 111209A shows a consistent behavior starting from about $T-T_0=30$ ks in the rest frame (Fig.\ref{f:igc}). Following \cite{Pisani2013}, we further make a blind search of the redshift by comparing the expected GRB 111209A X-ray luminosities it the burst were at redshift $z=$0.4,0.5,0.6,0.7 and 0.8, with the prototype light curve of GRB 090618, and we find that residuals are minimized for a redshift between 0.6 and 0.7, consistent with the spectroscopic redshift of z=0.677 of GRB 111209A.


\subsection{The optical temporal lag: evidence of two emitting regions?}
\label{s:lagdisc}

Though optical flare lags have been already observed in other LGRBs, the lag measured for GRB 111209A of $\sim245$ s in the rest frame is far more larger than what was reported earlier. For example, for GRB 081126 at redshift $2.8<z<3.8$ and with burst duration of $T_{90}\sim55$ s, a temporal lag of ($8.4\pm3.9$) s was measured for a prompt emission flare \citep{Klotz2009}.  For XRF 071031 at $z=2.692$ and with $T_{90}\sim180$ s an optical flare lag was measured to be 35 seconds in the burst frame \citep{Kruhler2009}. No temporal lag was found for a simultaneous prompt emission optical and gamma-ray flare detected in two very long GRB 110205A at $z=2.22$ \citep{Gendre2012} and GRB 100901A at $z=1.408$ \citep{Gorbovskoy2012}, with $T_{90}=260$ s and $T_{90}=439$ s respectively. A 2 s delay was observed between the gamma-ray and optical variable emission during the prompt emission of the very bright, ``naked eye" burst GRB 080319B at z=0.937 with $T_{90}>50$s \citep[e.g.][]{Beskin2010}.
All these temporal lags range below $\sim50$ seconds in the burst rest frame and  
there is no obvious connection with the burst duration that would explain the much longer delay of the ultra-long GRB 111209A. 

In general, flares observed in X-rays and gamma-rays are known to show faster rising, faster decaying profile, and earlier peak at high energies than at low energies due to a hard-to-soft spectral evolution \citep[e.g.][]{Margutti2011}. The optical lags observed in some GRBs may be an extension of this property in the optical energy domain. The large diversity on the optical temporal gaps observed among several LGRBs may be due to different dynamics of the synchrotron frequencies crossing the two energy domains \citep[e.g.][]{Kruhler2011}. 

In the framework of the internal shock model, another possible explanation of the optical temporal lag predicts that the optical counterpart of the gamma-ray flare is generated by synchrotron mechanism in a different, optically thin region of the ejecta, more distant from the inner, optically thick regions where gamma rays are generated. 
To reproduce two distinct emitting regions, one possible scenario is based on the assumption of a large neutron ejecta component: delayed flare optical counterpart can originate in the catching up of a late ejected proton shell with an earlier ejected neutron shell that has travelled far away from the central engine with negligible interactions with the ejecta, where electrons are produced through neutron $\beta$-decay \citep{Fan2009}. The predicted time delay between the optical and gamma-ray peaks is expected to be of the order of 2-3 seconds for $\Gamma\sim 300$ (from $\Delta t\sim1.1(1+z)300/\Gamma$ s, Fan et al. 2009), thus it can reproduce the large temporal lag observed for GRB 111209A only for a Lorentz factor of the neutron shell of a few. 
In an alternative scenario, a delayed optical emission can originate from ``residual" collisions at larger distance $R$ from the central engine \citep{Li2008}. Temporal delay of the order of a fraction of seconds is obtained for fiducial values of $R\sim10^{15}$ cm and $\Gamma\sim300$ from $\Delta t\sim R/2c\Gamma^2$. Large time delays as those observed in GRB 111209A, for which $R\sim1.5\times10^{13} \Gamma^2$ cm, could thus be achieved assuming low $\Gamma$ values or large radii. 


\subsection{The X-ray hard extra power law component origin}

In \S\ref{s:xmm} the temporally resolved spectral analysis of the XMM-Newton data suggest the presence of two spectral components at the end of the steep decay phase, where the hardest one dominates at high energies (see $\beta_1$ and $\beta_2$ in Tab.\ref{table:xmm} and Fig.\ref{f:xmm}).   
An intriguing explanation of the hard spectral component in the XMM-Newton spectra  can be found following the recent gamma-ray instruments findings in the GRB observational campaigns. Some bursts observed by Fermi in the Large Area Telescope (LAT) energy range (20 MeV - 300 GeV), as for instance GRB 090902B \citep{Abdo2009}, show the presence of a very hard power law spectral component in addition to the typical prompt emission spectrum that is usually fitted by a Band law or a cut-off power law. 

The power law extra-component is present during the prompt phase and it decays once the afterglow has started \citep[e.g.][]{Abdo2009}, and is very hard (e.g. an energy spectral index of $\beta_h=0.62 \pm 0.03$ for GRB 090510, \cite{Ackermann2010}). 
One may thus consider the extra-component seen in GRB 111209A as the ``soft tail" of this hard power law seen by the Fermi/LAT. The fact that we detect it only at the end of the steep decay phase is linked to its properties: the hard power law component intensity rises slowly during the prompt phase and it is typically detected when the X-ray prompt emission has just ended. At that point it decays following a power law (see e.g. Zhang et al. 2011 for a review). In our case, the soft tail of this component is too faint to be detected during the prompt emission while it emerges when the main X-ray prompt spectral continuum drops down to small flux levels at the end of the steep decay phase. By roughly modelling the temporal evolution of the hard component (see $F_2$ values in the temporal bins A, B and C in Table \ref{table:xmm}) with a power law decay, we could roughly measure a decay index of $\alpha_{h}=3.8^{+1.2}_{-3.0}$ that, despite the large uncertainties, is consistent with past measures (e.g. $\sim 0.8$ in the case of GRB 090902B, $\sim1.1$ for GRB 090510, see Ackermann et al. 2013) and it may be steeper than the afterglow decay that, at that epoch, is entering in the shallow decay phase thus preventing further detections at later times of the hard component.  

Ironically, one of the two GRBs for which the prompt extra spectral component has been unambiguously detected (GRB 090902B and GRB 090510) is a short GRB, that is, at the opposite end of the burst duration distribution range than the ultra-long GRB 111209A. The extra power law spectral feature was detected also for GRB 090510 that showed a burst duration of $\sim 20$ s.
 

\cite{Zhang2011} denote this extra power law as ``Component III", in addition to the Band model (``Component I") and a thermal component (``Component II"). Among the possible interpretations of Component III (see Zhang et al. 2011 and references therein), it has been proposed that it may originate from the Compton-upscattered emission of a simultaneous thermal emission in the MeV energy range (``Component II") that was observed in both GRB 090510 and 090902B. Although we detect a thermal emission for GRB 111209A, it was in the soft X-ray energy range and not simultaneous to the power law extra-component. 
An alternative scenario suggests ``Component III" to be emitted from another site using the classical emission mechanism for GRBs \citep{Zhang2011}. As discussed in \S\ref{s:lagdisc}, a two different emission sites scenario could stand for GRB 111209A. 


\subsection{The origin of the optical afterglow emission}
\label{s:sed}

In the following we attempt to interpret the origin of the early and late optical afterglow emission.  

For the early afterglow, given the large uncertainties affecting the rising phase of the optical on-set bump, we consider here only its decaying properties. We have seen that between 20 and 40 ks, the flux decays following a power law with and average index of $\alpha_{2}=1.6\pm0.1$ (Tab.\ref{tab:otlc}). In the same temporal interval, using {\it Swift}/UVOT data, we measure a spectral index of $\beta_{opt}=1.33\pm0.01$ (\S\ref{s:opt}). In the following, we explore three possible scenarios of this early emission. 

{\it Forward shock. } 
According to the synchrotron closure relationships between the temporal and spectral indices, the expected spectral slope during this phase is  $\beta=2\alpha/3=1.1\pm0.1$ or $\beta=(2\alpha+1)/3=1.4\pm0.1$ for $\nu_m<\nu<\nu_c$ or $\nu>\nu_c$, respectively where $\nu_c$ is the synchrotron cooling frequency and $\nu_m$ is the frequency at which the bulk of the electrons are radiating \citep{Sari1998}.  
The measured spectral index $\beta_{opt}$ is thus consistent within the uncertainties with the case $\nu_{opt}>\nu_c$. Alternatively, the case $\nu_{opt}<\nu_c$ may also be valid assuming a certain amount of dust extinction\footnote{We have considered the case of a slow cooling regime of the bulk of the electrons. In case of fast cooling, the two expected spectral indices are associated to the following frequency ranges $\nu_c<\nu<\nu_m$ and $\nu>\nu_m$ \citep{Sari1998}} . 


If FS is the correct interpretation, the bump at $T_0+10$ ks marks the fireball deceleration epoch. Assuming a constant density environment in the range $n=0.01-1$ cm$^{-3}$ and an energy conversion efficiency of $\eta=0.1-0.2$, we find $\Gamma_0$ of the order of $\sim50-120$ and a deceleration radius of $\sim 4-20\times10^{17}$ cm by simply equating the fireball energy with the swept-up mass energy assuming a spherical geometry \citep{Sari1998}.  The lack of evidence of a further steepening of the light curve before $\sim50$ ks, that is before the epoch when the rebrightening phase starts to dominate, imposes a lower limit to the jet opening angle. Assuming a range of density values between 0.01 and 1 cm$^{-3}$, we estimate that the jet half opening angle should be larger than $\sim$1 - 4 degrees\footnote{ We caution that the rebrightening at $\sim 1$ day after the trigger may indicate, among several scenarios, an energy injection or the emergence of an additional jet component. Therefore assuming the initial kinetic energy and the lower limit on the jet break time to estimate the jet opening angle of GRB 111209A may be an oversimplification.}


{\it Reverse shock.} An alternative interpretation of the early optical decay can be found in the reverse shock (RS), formed in the fireball impact with the surrounding medium. Indeed, RS emission is expected to produce early optical ``flashes" peaking at epochs comparable to the burst duration \citep{Kobayashi2000}.  This may be the case for the early optical emission of GRB 111209A, the decaying behavior of which starts roughly at the estimated end of the prompt emission, as expected. 
Given the ultra-long duration of the burst, it is likely that the so called ``thick shell" case applies for GRB 111209A.  
In the ``thick shell" case, the reverse shock has enough time to accelerate to relativistic velocities whithin the shell. The expected spectral behavior of the reverse shock follows the synchrotron emission prescriptions, as for the forward shock case. 
According to \cite{Kobayashi2000}, for $\nu_{opt}<\nu_c$, a flux decay is expected to have a power law index of  $\alpha=(73p+21)/96$, thus consistent with the measured decay index $\alpha_2$ (Tab.\ref{tab:otlc}) for an electron (power-law) energy distribution  index of $p\sim1.8$. 
We note that from the late afterglow radio-to-X-ray SED we inferr a higher value of $p$  (see \S\ref{s:radiosed}), thus if the RS is the correct interpretation, these results possibly indicate an evolution of the microphysical parameters between the early and the late afterglow phases.

For $\nu_{opt}>\nu_c$, the RS emission rises to a constant flux phase up to the end of the prompt duration and then it vanishes since no electrons are shocked after the RS has crossed the shell. Thus, if $\nu_{opt}>\nu_c$, the RS scenario cannot explain the observed smooth flux decay.


{\it Internal shock. }Another possible interpretation associates the steep optical decay to the high latitude emission from the internal shocks emission (IS) formed whithin the ejecta and responsible also of the X-ray and gamma-ray emission. However, in this case, the predicted decay rate is $\alpha=(2+\beta)\geq3$ \citep[e.g.][]{Kumar2000}, that is much steeper than the measured one, so we can exclude this scenario. In addition, the radius where IS takes place is typically smaller than the external shock; this makes unlikely an IS decay time scale of order of several tens of kiloseconds \citep[e.g.][]{Wu2013}.

Now we turn to interpret the re-brightening observed at $T_0+100$ ks. 
Before the re-brightening peak, the estimated mean rising index is $\alpha_{r,1}=-2.0\pm0.5$ (Tab.\ref{tab:otlc}) while the spectral index is $\beta_{opt}=1.0\pm0.1$ up to the re-brightening peak epoch (measured from NIR and optical data, \S \ref{s:opt}).    
The X-ray plateau is observed nearly simultaneously with the optical re-brightening. 
Several hypothesis have been made to interpret the late re-brightening observed in many LGRBs.

{\it Structured jet model.} A re-brightening feature can be reproduced by a two jet component scenario. One possible configuration predicts that the prompt and the early afterglow emission are produced by a fast, narrow jet while the late re-brightening is due to the deceleration process of a wide, slower jet, that may dominate the emission from the other jet at late times \citep[e.g.][]{Peng2005}. 
We have already seen that the early optical afterglow emission of GRB 111209A can be interpreted in the context of the FS emission, thus, in this scenario, with the FS formed in the impact between the narrow jet with the ISM.  However, if the rebrightening were produced by the FS of the wider jet, according to the FS modeling 	\citep{Sari1998}, we would expect a slower rising index and a much harder spectrum than what we measure during the rising phase. In addition we do not find marked evidence of spectral evolution near the peak epoch as expected from the crossing of the injection synchrotron frequency. 

An alternative scenario assumes that the two jets have different axis orientation with respect to the line of sight of the observer, where the off-axis one originates the rebrightening as an effect of the jet approaching the line of sight \citep[e.g.][]{Huang2004}. In this scenario a large variety of rising behaviours can be reproduced, depending on the angle formed between the line of sight and the jet axis. In this case, in order to dominate the on-axis jet emission at late times, the off-axis jet component should be more energetic. For GRB 111209A, for which $E_{iso}\geq10^{53}$ erg \citep[e.g.][]{Golenetskii2011}, the latter condition makes this scenario less favourable, although it cannot be excluded.   

{\it Late prompt model.} A large range of possible rebrightening morphologies can be reproduced also by the so called ``late prompt model" \citep[e.g.][]{Ghisellini2007}. In this model, the re-brightening feature is the emerging of the prolonged central engine emission from the decaying afterglow signal. This model predicts a strong spectral evolution during the rebrightening that is however not so evident for GRB 111209A (Fig.\ref{f:otlc}).  

{\it Density gradient. } One possibility invokes a density gradient in the circumburst environment if the condition $\nu<\nu_c$ is satisfied \citep[e.g.][]{Lazzati2002}. Since the first studies on this hypothesis, the vast   phenomenology observed so far have shown that additional requirements should be included in the density gradient scenario in order to reproduce the observations. In particular, \cite{Kong2010} were able to reproduce a large variety of late re-brightening morphologies by allowing the shock microphysics parameters to vary between an initial wind environment and the following shocked constant density environment.

{\it Delayed energy injection.} Another possible interpretation of the late re-brightening is associated with a late energy injection given by the delayed interaction of slow ejected shells with the fireball \citep[e.g.][]{Fan2006}. This model has been frequently invoked to explain the X-ray shallow phase observed in a large fraction of X-ray afterglows soon after the steep decay phase. 
Energy injection models do not predict any spectral variability and indeed during the plateau in X-rays of GRB 111209A, we could detect the presence of a spectral component with nearly constant spectral slope $\beta_X=1.7\pm0.1$ (Tab.\ref{table:xmm}). In this scenario, the late optical rebrightening is simply the optical counterpart of the X-ray emission, unless assuming that it is by chance simultaneous with the X-ray plateau.

\section{Summary and Conclusions}
\label{s:conclusions}
We have analyzed the extensive multi-wavelength data set of the ``ultra-long" GRB 111209A from the prompt emission to the late afterglow. During the prompt emission, at 2 ks from $T_0$ we measure an unprecedented large temporal lag of $\sim245$ s in the burst rest frame  between the optical and the high energy peak time of a pronounced flare. This lag may be the evidence of two distinct emission regions for the gamma-rays and optical observed radiation, although other scenarios cannot be excluded. 

Separate emission regions may be also supported by the evidence of an extra hard power law component in the X-ray spectrum at the end of the prompt emission if interpreted as the soft tail of the hard power-law component observed in few cases by gamma-ray instruments at the prompt-to-afterglow transition phase (e.g. Zhang et al. 2011). 

Assuming a common origin of the optical and the high energy prompt photons, a non negligible amount of dust extinction should be invoked, in analogy with other very long GRBs for which simultaneous optical to gamma-ray prompt emission data were available (e.g. GRB 100901A \cite{Gorbovskoy2012}) and with two very dark GRBs with prompt duration of about 800s \citep{Zauderer2013}. The afterglow data analysis, however, does not confirm the presence of dust, possibly indicating that the intense UV and X-ray flux from the GRB have partially destroyed the dust along the line of sight in the host galaxy. 

The presence of dust is consistent with the findings of a subsolar but not exceptionally low metallicity in the host galaxy of GRB 111209A and is at odds with the very low metallicity envirnoment invoked for a blue supergiant progenitor in our Paper I. A possible solution can be found by invoking a binary system formation channel of the blue supergiant, as proposed by \cite{Podsiadlowski1992}. The evolution of a binary system of two massive stars or a massive star and a small mass companion has often been invoked for LGRBs to confront the increasing evidence of a significant fraction of LGRBs in high metal-content host galaxies (see e.g. Levesque 2013 for a recent review). 
In addition, binary systems are at the basis of the IGC model that provides a unified view of the phenomenology observed in both the prompt and the afterglow observed in some LGRBs \citep{Ruffini2001}. 

Despite the exceptional longevity of GRB 111209A, its afterglow is not dissimilar from other normally long GRBs in terms of optical and X-ray afterglow light curves and spectra. Indeed, the prompt to afterglow transition in both X-rays and optical wavelengths does not differ from several other LGRBs studied in the past, showing an initial flux decay with different decay indices in the two energy ranges and followed by an optical re-brightening nearly simultaneous to an X-ray plateau. 

The spectral and temporal information available during the early optical afterglow decay prevent us to disentangle among a forward shock or a reverse shock origin, while enable to exclude an internal shock origin. In the FS case, a Lorentz factor of $\sim50-120$ and a deceleration radius of $\sim 4-5 \times 10^{17}$ cm are inferred. 

The late afterglow tentatively favours the presence of a prolonged energy injection from the central engine that, if true, should be active up to 70-80 ks after the trigger, that is, for a dozen of hours in the rest frame.  However, the late rebrightening can be explained also through a density gradient, as expected in the likely complex environment surrounding the progenitor of GRB 111209A, or by a structured jet scenario. The radio to X-ray broad band SED at 5 days after the trigger is consistent with the synchrotron model for a fireball expanding in an ISM with density $\sim0.1$ cm$^{-3}$ and a predicted large half opening angle of the jet ($\sim23$ degrees).

Finally, we note that the existence of a bright optical emission associated with an ultra-long GRB with flux level above 0.1-1 mJy up to several hours after the trigger may increase the orphan afterglow detection probability, a still unsolved fundamental issue that would definitively provide access to the jetted nature of these events \citep[e.g.][]{Atteia2013}. 


\acknowledgments
We thank the anonymous referee for useful comments and suggestions that have lead to an overall improvement of the manuscript.  We aknowledge A. D. Kann for useful discussion. 
This work made use of data supplied by the UK {\it Swift} Science Data Centre at the University of Leicester, it is a result of the FIGARO collaboration, and was supported by the French Programme National des Hautes Energies and by the ASI grant I/009/10/0. We acknowledge the use of public data from the {\it Swift} data archive. TAROT has been built with the support of the CNRS-INSU. We thank the technical support of the XMM-Newton staff, in particular N. Loiseau and N. Schartel. SO and MDP acknowledge funding from the UK Space Agency.

{\it Facilities:} \facility{Swift}, \facility{XMM-Newton}, \facility{TAROT-Calern}, \facility{TAROT-La Silla}.

\clearpage

\end{document}